\documentclass[%
 reprint,
 onecolumn,
 amsmath,
 amssymb,
 aps,
 unsortedaddress
]{revtex4-2}

\usepackage{graphicx}
\usepackage{bm}
\usepackage{subcaption}
\usepackage{makecell}
\usepackage{multirow}
\usepackage{amsmath,amssymb,amsthm,easybmat,verbatim}
\usepackage{algorithm}

\usepackage{algpseudocode}
\usepackage{caption}
\usepackage{xcolor}
\usepackage{listings}
\usepackage{hyperref}
\usepackage{float}
\hypersetup{
    colorlinks,
    linkcolor=blue,
    citecolor=blue,
    urlcolor=blue
}

 \theoremstyle{remark}

\usepackage{orcidlink}

\algnewcommand{\To}{\textbf{To }}
\algnewcommand\Input{\item[\textbf{Input:}]}%
\algnewcommand\Output{\item[\textbf{Output:}]}%
\algnewcommand\Func{\item[\textbf{Function:}]}
\newlength\myindent
\usepackage{physics}

\usepackage{ulem}

\usepackage{colortbl}
\usepackage{tikz}
\usetikzlibrary{positioning,fit,backgrounds}

\usepackage{xparse}

\NewDocumentCommand{\phaseone}{ m O{0} O{1} o }{%
  \begin{tikzpicture}[baseline={(0,-0.25)}]
    \begin{scope}[local bounding box=#1]
      \draw (0,0) node (0) [main node] {#2};
      \draw (1,0) node (1) [main node] {#3};
      \IfValueT{#4}{%
        \draw (.5,-.5) node [label node] {$t = #4$};
      }%
      \path (1) edge [in=30, out=60, loop, above right, inner sep=0.02cm] (1);
    \end{scope}
  \end{tikzpicture}%
}

\NewDocumentCommand{\phasezero}{ m O{0} O{1} o }{%
  \begin{tikzpicture}[baseline={(0,-0.25)}]
    \begin{scope}[local bounding box=#1]
      \draw (0,0) node (0) [main node] {#2};
      \draw (1,0) node (1) [main node] {#3};
      \IfValueT{#4}{%
        \draw (.5,-.5) node [label node] {$t = #4$};
      }%
      \path (0) edge [in=30, out=60, loop, above right, inner sep=0.02cm] (0);
    \end{scope}
  \end{tikzpicture}%
}

\NewDocumentCommand{\phaseboth}{ m O{0} O{1} o }{%
  \begin{tikzpicture}[baseline={(0,-0.25)}]
    \begin{scope}[local bounding box=#1]
      \draw (0,0) node (0) [main node] {#2};
      \draw (1,0) node (1) [main node] {#3};
      \IfValueT{#4}{%
        \draw (.5,-.5) node [label node] {$t = #4$};
      }%
      \path (0) edge [in=30, out=60, loop, above right, inner sep=0.02cm] (0);
      \path (1) edge [in=30, out=60, loop, above right, inner sep=0.02cm] (1);
    \end{scope}
  \end{tikzpicture}%
}

\NewDocumentCommand{\edgehor}{ m O{0} O{1} o }{%
  \begin{tikzpicture}[baseline={(0,-0.25)}]
    \begin{scope}[local bounding box=#1]
      \draw (0,0) node (0) [main node] {#2};
      \draw (1,0) node (1) [main node] {#3};
      \IfValueT{#4}{%
        \draw (.5,-.5) node [label node] {$t = #4$};
      }%
      \path (0) edge [inner sep=0.02cm] (1);
    \end{scope}
  \end{tikzpicture}%
}

\NewExpandableDocumentCommand{\zhZ}{O{}t*t-m}{
  \node[zx main node, zhZ4={#1}{\IfBooleanTF{#3}{-}{}}{\IfBooleanTF{#2}{*}{}}{#4}] {}; %
}
\NewExpandableDocumentCommand{\zhX}{O{}t*t-m}{
  \node[zx main node, zhX4={#1}{\IfBooleanTF{#3}{-}{}}{\IfBooleanTF{#2}{*}{}}{#4}] {}; %
}
\NewExpandableDocumentCommand{\zhH}{O{}t*t-m}{
  \node[zx main node, zhZ4={#1}{\IfBooleanTF{#3}{-}{}}{\IfBooleanTF{#2}{*}{}}{#4}, rectangle] {}; %
}

\usepackage{booktabs}

\makeatletter
\expandafter\edef\csname ftype@algorithm\endcsname{\the\value{float@type}}
\makeatother

\begin{document}

\preprint{APS/123-QED}

\title{Quantum-informed graph decomposition}

\author{Matthew Brunet}
\author{Leah Prepchuk}
\author{Rebekah Herrman
}
\email{rherrma2@utk.edu}
\affiliation{Department of Industrial and Systems Engineering, University of Tennessee Knoxville, USA}


\date{\today}

\begin{abstract}
The Quantum Approximate Optimization Algorithm (QAOA) is a near-term quantum algorithm for solving combinatorial optimization problems, but its accuracy and resource requirements may scale poorly with problem size or structure, reducing its utility in practice. Approaches such as the Graph Decomposition for Optimization Algorithm (GDOA) seek to address this by iteratively reducing an input problem until it reaches a size where it can be solved using QAOA while maintaining the same optimal solution. This algorithm, however, requires solving integer programs that have exponentially many variables, and can result in suboptimal solutions, as the integer programs may contain slack terms. In this work, we propose a heuristic called the Quantum Enhanced Graph Decomposition (QEGD) algorithm. Its novelty lies in its use of QAOA as a subroutine for sparsifying a graph through local vertex-merging.  The algorithm, when used to sparsify MaxCut problems that are then solved using QAOA, achieves higher approximation ratios across all tested graphs than GDOA for random 5-regular and 6-regular graphs. We then compare the time-to-solution and approximation ratio of QEGD to other QAOA variants a collection of Erd\"{o}s-R\'{e}nyi graphs. QEGD has higher approximation ratios on average than QAOA$^2$ and ParaQAOA on all tested graphs. It also results in higher average approximation ratios than DC-QAOA on sparse Erd\"{o}s-R\'{e}nyi graphs and comparable average approximation ratio on denser Erd\"{o}s-R\'{e}nyi graphs. However, its average wall-clock time is between 0.4\% and 5\% of the average DC-QAOA wall-clock time.
\end{abstract}

\maketitle

\section{Introduction}\label{sec:intro}
 The Quantum Approximate Optimization Algorithm (QAOA) is a hybrid variational algorithm used for solving combinatorial optimization problems, such as the MaxCut problem \cite{farhi2014quantum}. The algorithm requires a quantum circuit that depends on classical parameters, commonly assigned using a classical subroutine, to explore the solution space. Recent work has shown that the algorithm may have advantages over classical algorithms when solving combinatorial optimization problems. For example, Shaydulin et al. demonstrated numerical evidence of a quantum speedup in the low autocorrelation binary sequence (LABS) problem, outperforming the state-of-the-art branch and bound techniques for that problem \cite{LABSScaling2024}. Empirical studies show that QAOA can provide a speedup in complexity over the best classical solvers for the 8-SAT problem \cite{ksat2024}. When combined with a linear ramp parameter schedule, there exists a scaling advantage in time-to-solution over classical algorithms for the weighted Max-Cut problem \cite{montanez2025, Blekos2023ARO}. 
 
 As problem size grows, deeper and wider quantum circuits are required to converge to an optimal solution, introducing more noise and thus more error. These issues have been the impetus behind many variations of QAOA, such as QAOA$+$, multi-angle QAOA (MA-QAOA), and Grover-Mixer QAOA (GM-QAOA), all of which attempt to either improve solution quality, decrease circuit depth by changing the algorithm ansatz, or allowing for additional parameters, increasing expressivity \cite{herrman2022multiangle,bartschi2020grover, Hadfield_2019}. Additionally, recent QAOA research has also examined classical pre- and post-processing techniques, such as graph partitioning, algebraic problem reformulation, or machine learning, to improve QAOA performance \cite{ brunet2025exclusive, mlqaoa2020}. For example, QAOA-in-QAOA (QAOA$^2$)  partitions combinatorial optimization problem graphs into disjoint subproblems using community detection methods, solves the subproblems in parallel using QAOA, then merges them together and performs QAOA once more \cite{zhou2023qaoa}. The Divide-and-Conquer QAOA (DC-QAOA) parses a graph with a recursive, large graph partitioning algorithm, solves the sub-problems in parallel, then produces the final solution with an quantum state reconstruction technique based on bit-string agreement \cite{li2022large}. ParaQAOA employs a connectivity preserving partition algorithm to divide a problem graph into subproblems, solves the subproblems in parallel with QAOA, then reconstructs the final solution with level-aware parallel merging\cite{huang2026paraqaoaefficientparalleldivideandconquer}.

In a similar vein is the Graph Decomposition for Optimization Algorithm (GDOA)\cite{ponce2025graph}. The goal of this algorithm is to transform one MaxCut instance into a new MaxCut instance that has fewer vertices and the same optimal objective value as the original MaxCut instance. It achieves this through an iterative minimum vertex separator reweighting scheme which is guaranteed to remove at least one vertex per round. However, the reliance on solving integer programs results in not only a subroutine that grows exponentially in the size of the smallest cut set of the graph, but also introduces slack variables, and thus a potential for inexact solutions that may result in a suboptimal objective value.

In contrast to using classical methods to improve QAOA, quantum-informed recursive algorithms (QIRAs) utilize insight from one iteration of QAOA to inform classical optimization algorithms. Recursive QAOA (RQAOA) pioneered this format. The algorithm implements greedy problem reductions using QAOA expectation values, as delineated in Section II B \cite{rqaoaOriginal}. Building on this, Finzgar et. al. proposed a Quantum Informed Recursive Optimization (QIRO) algorithm uses QAOA expectation values to eliminate variables in Maximum Independent Set (MIS) and Max-SAT problems \cite{PRXQuantum1}.

In this work, we propose a Quantum-Enhanced Decomposition Algorithm (QEGD) which does not rely on parallel architecture, exponential solution reconstruction, nor complex graph partitioning. Instead, QEGD takes influence from recent QIRAs. The goal of the algorithm is to use QAOA expectation values to reduce the minimum cardinality cut set of a problem graph whenever it is greater than three, as GDOA can guarantee optimality when the cut set has cardinality at most three. Then, one can apply GDOA to further reduce the problem graph. We compare the approximation ratios and time-to-solution of QEGD combined with GDOA against QAOA$^2$, DC-QAOA, and ParaQAOA when solving MaxCut on Erd\"os- R\`enyi random graphs with between 20 and 30 vertices and edges probabilities between 0.1 and 0.8. We find that QEGD combined with GDOA  outperforms both QAOA$^2$ and ParaQAOA in approximation ratio for sparse ER graphs by over 12\% and 15\% respectively, while at the same time maintaining comparable speeds to QAOA$^2$. Furthermore, the runtime of QEGD is on the order of a dozen seconds for the test graphs, comparable to QAOA$^2$ and one-hundredth of the runtime of DC-QAOA on the same graphs. Not only that, but QEGD obtains a 15\% higher approximation ratio than GDOA alone on 6-regular 100-node graphs, and a 10\% higher approximation ratio on 5-regular graphs. Since GDOA only requires QAOA expectation values on small subgraphs of an initial problem graph, the quantum overhead is minimal. Additionally, since it only requires one iteration of QAOA to calculate expectation values, QAOA data can be calculated via closed form equations \cite{hadfield2018quantum}, obviating the need to run QAOA on hardware. 

This work is organized as follows. In Sec.~\ref{sec:background}, we introduce QAOA notation, the MaxCut problem, iterative quantum algorithms, and GDOA. We then introduce QEGD in Sec.~\ref{sec:qegd}. We then compare QEGD to other quantum algorithms in Sec.~\ref{sec:results}. Finally, we close with a discussion in Sec.~\ref{sec:concs}.
 
\section{Background}\label{sec:background}
In this section, we introduce QAOA, distributed QAOA variants, the MaxCut problem, QIRAs, and the GDOA.

\subsection{QAOA} 

QAOA is variational quantum algorithm consisting of two unitaries, the cost unitary, $U_C(C,\gamma) = e^{-iC\gamma}$, and the mixing unitary, $U_B(B,\beta) = e^{-iB\beta}$, applied in an alternating fashion to an initial quantum state $\ket{s}$. The Hamiltonian $C$ is a diagonal matrix that encodes the combinatorial optimization problem being solved. $B$ is a matrix with non-diagonal entries that is used to explore the solution space, and in order to guarantee QAOA convergence, $\ket{s}$ should be an eigenstate of $B$. Often $B$ is the transverse field mixer $\sum_i{X_i}$ where $X_i$ is the Pauli-x matrix acting on qubit $i$, and $\ket{s} = \ket{+}^{\otimes n}$. The parameters $\gamma, \beta$ (often called angles) are typically found via classical optimization routines like COBYLA or BFGS \cite{QiskitAer}. A layer of QAOA is defined as one application of $U_B(B,\beta_i)U_C(C,\gamma_i)$ to the initial state $\ket{s}$, and we let $p$ be the number of layers. The number of parameters required to implement $p$ layers of QAOA is $2p$. 
The primary success metric of QAOA is the approximation ratio (AR), defined as
\begin{align*}
    \frac{\langle C \rangle}{C_{max}}
\end{align*}
where $C_{max}$ is the largest  objective value of the optimization problem, and $\langle C \rangle$ is the expected value of the QAOA cost Hamiltonian. The AR is a real number between 0 and 1, where an AR closer to 1 indicates the measured QAOA solution is closer to the optimal solution. 

\subsection{Distributed QAOA}
Distributed QAOA methods use divide-and-conquer-style approaches to solve combinatorial optimization graph problems at scale. Algorithms such as DC-QAOA and ParaQAOA are constructed using graph partition subroutines, and are designed such that one implements QAOA on smaller subproblems, so each QAOA circuit that solves a subproblem requires fewer qubits. A keystone part of all distributed QAOA algorithms is their reconstruction techniques, which require QAOA-solved subproblems as input and reconstruct a solution to the original combinatorial optimization problem. DC-QAOA uses a bitstring distribution frequency ranking system to recombine subproblems, whereas ParaQAOA takes the most optimal bitstring solutions from subgraph. These approaches, though, have drawbacks. ParaQAOA's parallelization relies on specific GPU architectures and it offers lower approximation ratios on non-regular graphs \cite{huang2026paraqaoaefficientparalleldivideandconquer}. DC-QAOA relies heavily on expensive partitioning and extra sampling for reconstruction fidelity, taking thousands of seconds for even 50 node graphs \cite{huang2026paraqaoaefficientparalleldivideandconquer, li2022large}.

\subsection{MaxCut}
Given a graph $G = (V,E)$, the objective of the MaxCut problem is to partition $V$ into two sets which maximize the number of edges that have endpoints in each set. If each edge $ij \in E$ has an associated weight, $w_{ij}$, the QAOA cost matrix $C$ of a MaxCut instance with weighted edges is 
\begin{align}
    C = \frac{1}{2}\sum_{(i,j) \in E}w_{ij}(1-Z_iZ_j)
\end{align}
where $Z_i$ and $Z_j$ are the Pauli-Z matrices acting on qubits $i$ and $j$, respectively. Importantly, \cite{wang2018quantum} derived a closed formula to calculate $\langle C_{ij}\rangle$ for $p=1$ unweighted MaxCut, which was then extended in \cite{Ozaeta_2022} to the weighted closed form. For adjacent vertices $i,j$, let $N(v)$ denote the neighborhood of $v$ and let $D = N(i)\cap N(j)$ be the set of their common neighbors. The weighted expectation can then be written as
\begin{align}\label{eq:p1}
    \langle C_{ij}\rangle
    &= \frac{w_{ij}}{2}
    + \frac{w_{ij}}{4}\sin(4\beta)\sin(w_{ij}\gamma)
      \left[\prod_{l\in N(i)\setminus\{j\}}\cos(w_{il}\gamma)
          + \prod_{k\in N(j)\setminus\{i\}}\cos(w_{jk}\gamma)\right] \nonumber\\
    &\quad - \frac{w_{ij}}{4}\sin^2(2\beta)
      \prod_{l\in N(i)\setminus(D\cup\{j\})}\cos(w_{il}\gamma)
      \prod_{k\in N(j)\setminus(D\cup\{i\})}\cos(w_{jk}\gamma)
      \left[\prod_{s\in D}\cos\big((w_{is}-w_{js})\gamma\big)
          - \prod_{s\in D}\cos\big((w_{is}+w_{js})\gamma\big)\right].
\end{align}
Notably, the QAOA expectation values after a single iteration can be computed analytically and without sampling. Algorithms in the following sections use the expected value of qubits, $\langle Z_iZ_j \rangle$, so we define it here as
\begin{align}\label{eq:EXP1}
    \langle Z_iZ_j\rangle
    &= 1 - \frac{2}{w_{ij}}\langle C_{ij}\rangle \nonumber\\
    &= \frac{1}{2}\sin^2(2\beta)
      \prod_{l\in N(i)\setminus(D\cup\{j\})}\cos(w_{il}\gamma)
      \prod_{k\in N(j)\setminus(D\cup\{i\})}\cos(w_{jk}\gamma)
      \left[\prod_{s\in D}\cos\big((w_{is}-w_{js})\gamma\big)
          - \prod_{s\in D}\cos\big((w_{is}+w_{js})\gamma\big)\right] \nonumber\\
    &\quad - \frac{1}{2}\sin(4\beta)\sin(w_{ij}\gamma)
      \left[\prod_{l\in N(i)\setminus\{j\}}\cos(w_{il}\gamma)
          + \prod_{k\in N(j)\setminus\{i\}}\cos(w_{jk}\gamma)\right].
\end{align}

\subsection{IQAs}
 Many recently developed algorithms have been developed as ``iterative quantum algorithms" (IQAs). Instead of executing a quantum algorithm on a problem, IQAs use a quantum subroutine in conjunction with classical techniques to progressively build a solution to the target CO problem. In IQAs, typically one applies one iteration of a quantum algorithm, such as QAOA, to a target problem. One then measures the expected values of the qubits or correlations between pairs of qubits, and uses this information to guide a classical graph decomposition heuristic to reduce the problem input size.
 
 One such iterative quantum algorithm is the Recursive QAOA (RQAOA)\cite{rqaoaOriginal}. When solving a MaxCut problem with RQAOA, one executes a single layer QAOA circuit, and measures the expected value of each qubit and correlations between pairs of qubits. These quantities can be calculated as \\
\begin{align}
    \langle Z_i \rangle & \approx \frac{1}{N}\sum_{N}s_i=P(z_i = 0)-P(z_i = 1)\label{eq:SINGLEEXP}\\
    \langle Z_iZ_j \rangle& \approx \frac{1}{N}\sum_{N}s_is_j = P(z_i =z_j)-P(z_i \neq z_j)
\end{align}
where $z_i$ is the $i$th qubit measured in the computational basis, and $s_i = (-1)^{z_i}$. As Eq.~\eqref{eq:SINGLEEXP} indicates, the single qubit expectation values are obtained by summing the observed values of said qubit across all measurements, and normalizing. For two-qubit correlations, the product of both observed qubit states is calculated for each measurement before being summed and normalized. After obtaining the QAOA expectation values, one performs a classical reduction step, which consists of substituting variables in the cost Hamiltonian. If $\langle Z_iZ_j\rangle > 0$, then the substitution $Z_i \rightarrow Z_j$ is made, and if $\langle Z_iZ_j\rangle < 0$, then substitution $-Z_i \rightarrow Z_j$ is made. These substitutions ensure that at least one variable is eliminated from the Hamiltonian, yielding a problem that requires fewer qubits to solve. This process repeats until only a predetermined number of variables remain, and then the reduced problem is solved using either a classical or quantum solver. The substitutions from earlier iterations are tabulated so that after solving the final instance, reverse substitutions are made to rebuild the full solution of the original problem instance. The RQAOA methodology of reducing a problem instance based on QAOA correlation values has been broadly applied in what are called quantum informed recursive algorithms (QIRA), which are a subset of IQAs. Recent QIRA research has focused on developing custom problem reduction schemes based on the class of problem being solved, such as \cite{PRXQuantum1, Wybo} which implement a distinct variable elimination rule for the Maximum Independent Set problem.

Many state-of-the-art QIRAs follow a ${``}$top-down${''}$ or $``$big-to-small$''$ algorithmic design pattern. Specifically, many of these algorithms glean expectation values from the entire problem instance before reducing the problem size \cite{PRXQuantum1, Wybo, rqaoaOriginal}. Functionally, the $``$big-to-small$''$ approach poses issues for scalability, both in quantum circuit depth and width \cite{farhi2020quantum}. These approaches do in many ways effectively combat the performance issues caused by the locality of QAOA, such as RQAOA perfectly solving the ring of disagrees at $p=1$, whereas QAOA can only achieve the approximation ratio $(2p+1) / (2p+2)$ \cite{rqaoaOriginal}. Especially for constraint enforced problems such as MIS, QIRAs achieve approximation ratios on par with supurb classical solvers, unlike the approximation ratios that QAOA can achieve on these problems \cite{Wybo}. Yet, they do so at the cost of potentially increased quantum resources. Additionally, RQAOA and its variants are characterized by their selection of the sole greatest expectation value to use for reduction, constraining each iteration to an elimination of only one variable.

\subsection{Graph Decomposition as an Optimization Algorithm (GDOA)}
\begin{figure}
    \centering
    \includegraphics[width=0.8\linewidth]{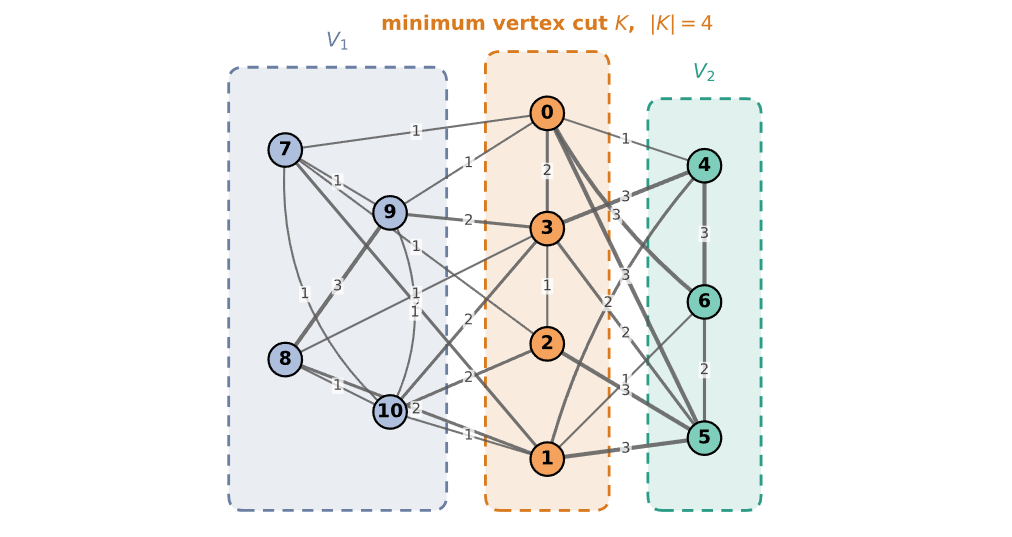}
    \caption{An example graph showing the minimum vertex cut partition}
    \label{fig:exGraph}
\end{figure}

Consider a graph $G=(V, E)$. The smallest set $K \subset V$ such that $G \setminus K $ is disconnected is called the minimum vertex cut of $G$, an example of which is shown in orange in Fig. \ref{fig:exGraph}. Removing the minimum vertex cut, $K$, of a graph yields at least two disconnected subgraphs, the vertices of which are denoted $V_1$ and $V_2$, such that there exists no edge connecting any vertex in $V_1$ to any vertex in $V_2$.  
Recent work by Ponce et al. details a classical algorithm which leverages a minimum cardinality vertex cut set $K$ of a graph to contract the graph into a graph with fewer vertices and weighted edges \cite{ponce2025graph}. The MaxCut problem instance of the contracted graph maintains an optimal objective value bounded above by the initial graph's maximum cut value \cite{ponce2025graph}. This is achieved through an iterative edge reweighting algorithm that removes one of the disjoint subgraphs induced by $K$ each iteration, and adds new, weighted edges between the nodes of $K$. These new edge weights are found using linear programming methods.
This process results in a graph that contains fewer vertices, but may potentially contain more edges. This trade-off allows for one to use fewer qubits to solve a MaxCut instance, however the circuit that solves the instance may require more gates, as each edge in the resulting graph results in two-qubit terms in the cost Hamiltonian. The authors showed that graphs up to 100 nodes in size could be solved with just $\approx 9.3$ qubits on average after decomposition, which is an approximately  $\approx 90\%$ reduction in qubits.  

However, there are pitfalls within the aforementioned decomposition algorithm. The most stringent is that in any given iteration, the reweighting sub-routine must solve $O(2^{|K|})$ linear programs of size $|V_2|$, where $V_2$ is assumed to be the component of $G\setminus K$ with smallest cardinality. The authors note that the graph that results from the decomposition algorithm has the same objective as the original graph if $|K| \leq 3$, however, there are error terms in the edge reweighting subroutine for larger $K$. 
Thus, it is desirable to maintain a consistently small $K$ to capitalize on their algorithm.

\section{The Quantum Enhanced Graph Decomposition Algorithm}\label{sec:qegd}
The Quantum-Enhanced Graph Decomposition Algorithm (QEGD) is a heuristic that is used to decompose a problem instance $G$ into a smaller instance with fewer vertices. The QEGD algorithm takes as input 1) a graph, $G = (V,E)$ with nodes $i \in V$. Each node is assigned a variable $x_i$ in the linear programming formulation of the combinatorial optimization problem, 2) a constant $M$ which is the target size of the reduced problem, 3) and a constant $-1 \leq \tau \leq 1$ which serves as a threshold for the merge subroutine. Then, while the $|V| > M$, the algorithm iterates over the following routine: it identifies a minimal cardinality cutset $K$, and vertex sets $V_1$, and $V_2$ which are edge-disjoint subgraphs of $G \setminus K$. An example of such a partition is outlined in Fig. \ref{fig:exGraph}, where the removal of $K$ disconnects the graph into at least two components, $V_1,V_2,\dots, V_m$ where $2 \leq m <|V_G|$. Without loss of generality, we let $m=2$ throughout this work, noting that when $m>2$, we can still treat the union of $m-1$ edge-disjoint subgraphs as a single disconnected subgraph. It should be noted that there is some component $V_i$ that has the smallest cardinality vertex set out of all components. If $|K| \leq 3$, then we use GDOA to reduce the graph, as it results in a provably optimal problem reweighting. Then $K \times K$ is reweighted, and then $V_2$ is deleted from the graph, ending this iteration. If, however, $|K| > 3$, the merge heuristic is triggered.

\subsection{Heuristic merge}
In this subroutine, we first run one iteration of QAOA on $K \cup V_2$, and collect the expectation values, $\langle Z_iZ_j \rangle$, $i \neq j \in K$. This pairwise expectation value ranges from -1 to +1, where values closer to -1 indicate that $x_i \neq x_j$, within the optimal MaxCut solution, and values closer to +1 indicate that $x_i = x_j$.
The magnitude of the correlation value itself indicate the confidence of said relationships. Notably, the algorithm does not require quantum hardware or simulation, as there exists closed formulas for the expectation values of one and two-qubit interactions using QAOA. The remainder of this section describes the algorithm in more detail.

\begin{figure}
    \centering
    \includegraphics[width=1.05\linewidth]{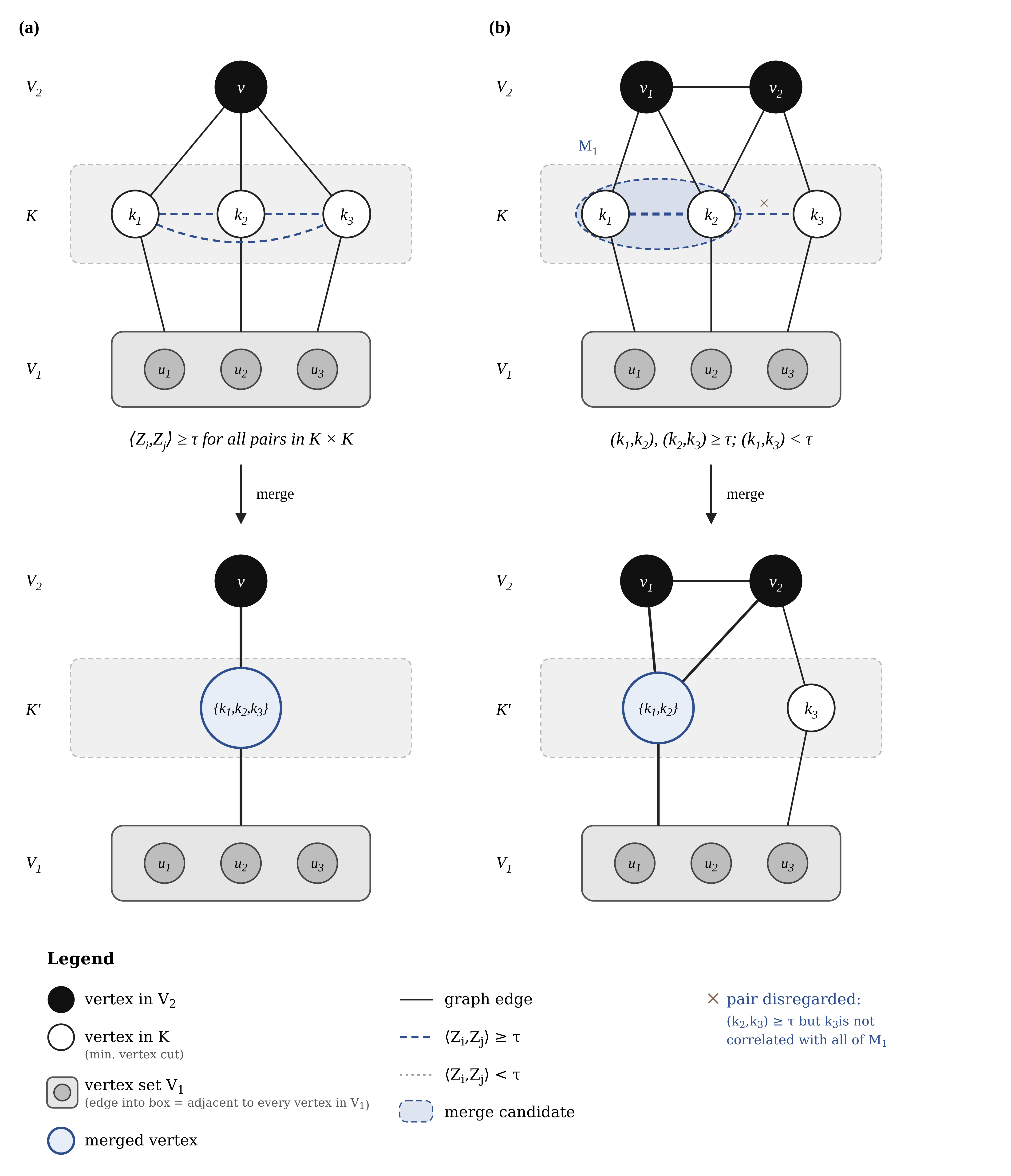}
    \caption{(a) Pairwise $\langle Z_iZ_j\rangle$ is obtained for all merge candidates $\{k_1,k_2,k_3\}$ and all correlations surpass $\tau$, so they are all merged into one supernode. (b) Only two of the three pairwise correlations exceed $\tau$, so only $\{k_1,k_2\}$ are merged. }
    \label{fig:mergeGraphic}
\end{figure}

The constant $\tau$ is used as a threshold, $\langle Z_iZ_j \rangle \geq \tau$, to guide a group vertex merge routine as follows.
All correlations greater than or equal to $\tau$ are sorted as vertex variable pairs from largest to smallest. The first pair of vertex variables $x_ix_j$ is placed into a list as a confirmed merge candidate. For subsequent pairs, $x_k x_\ell$, we check if either $k \in \{i,j \}$ or $\ell \in \{i,j \}$. If yes, (without loss of generality $k=i$), then we check all correlations $\langle Z_\ell Z_m \rangle$ for all $m \notin \{i,j,k,\ell \}$. If $\langle Z_\ell Z_m \rangle \geq \tau$ for all such pairs, we add $\ell$ to the merge candidate list. If no, then neither vertex in the pair is added to the confirmed candidate list and is instead added to a separate list. This process is repeated until all candidate pairs are exhausted. This clique merge constraint is outlined in Fig. \ref{fig:mergeGraphic}.\

\begin{figure}
    \centering    \includegraphics[width=1.0\linewidth]{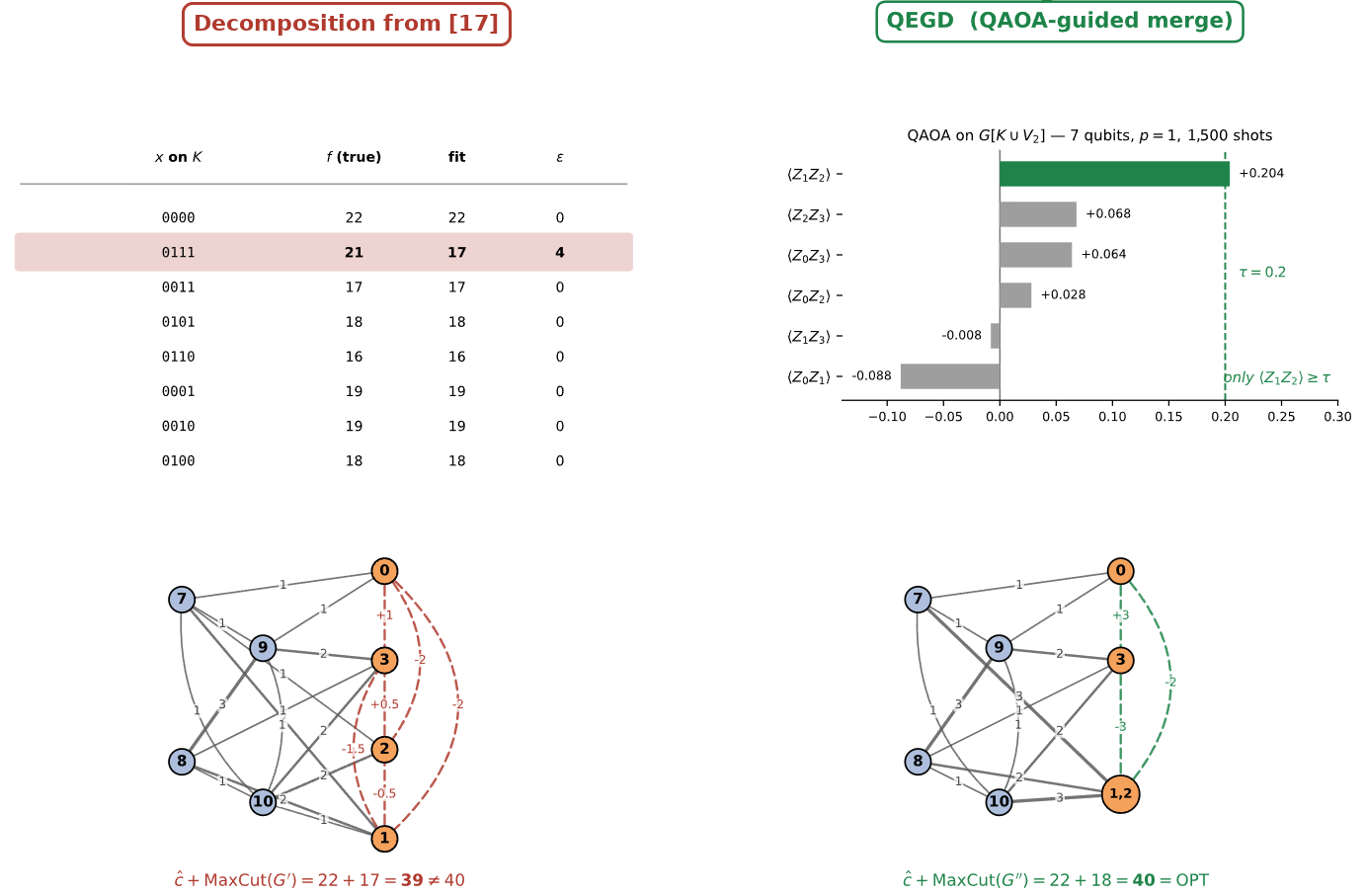}
    \caption{An example of reweighting where GDOA (left) provides a suboptimal solution and and QEGD (right) provides an optimal solution.}
    \label{fig:comp}
\end{figure}

The merge itself occurs by deleting each vertex in a confirmed merge candidate, then creating a new ``super-node'', $x_r$. Edges are added between the new super-node and all vertices that were in the neighborhood of the merged vertices. 
This newly reduced $K' \cup V_2$ is then passed to GDOA for reweighting if $|K'| \leq 3$. Otherwise, QEGD is performed again until $|K'| \leq 3$ or no $\langle Z_iZ_j\rangle \geq \tau$. The reduced graph $G'$ with nodes $K' \cup V_1$ becomes the input for the next iteration. The algorithm terminates when $|V| \leq M$, and the reduced graph is solved using either a classical or quantum algorithm. Once solved, all vertices that are in each super-node are assigned the same value as that the super-node. The cut value is equal to the objective of the reduced graph plus the cardinality of the super-node. The tabulated merges are then used to expand and assign sets to the nodes of the original graph. Pseudocode for QEGD is found in Algorithm~\ref{alg:pseudo}. Fig.~\ref{fig:comp} illustrates that there exist instances where one can use QEGD as a subroutine and achieve a larger size cut than GDOA.

\begin{algorithm}
\caption{QEGD}
\label{alg:pseudo}

    \begin{algorithmic}[1]
    \Input{undirected graph $G^0 = (V^0, E^0, J^0 )$, base case terminator $M$}
    \Procedure{Decomp\_Merge}{$G^0,M$}
    \State{initialize $i = 0, \;c^0 = 0, \;G^i = G^0, \;K^0 = S=\emptyset,  \;T:S\rightarrow V_s\subseteq V^0$ }
    \While{$|V^i| > M$}
    \State {$K^i =$ minimum\_cut\_set($G^i)$, $V_1^i,V_2^i = G^i \;\setminus K^i$} \Comment{$|V^i_1| \geq |V^i_2|$}
    \If{$|K^i| > 3$}
    \Procedure{Merge\_and\_Update}{$G^i,K^i,V_2^i$}
    \State{\textbf{return} $L,\hat{K}^i ,\hat{G}^i$}\Comment{$L$ maps new super-nodes to subsets of $K^i$}
    \EndProcedure
    \Procedure{Update\_Merge\_Table}{$L,t$}
    \State{Update global map $t$ to show which super-nodes go to which nodes in $V^0$}
    \EndProcedure
    \State{$G^i = \hat{G_i}$}
    \State{$K^i = \hat{K}^i$}
    \EndIf
    \State{$H^i = G^i[K^i\cup V^i_2]$}
    \Procedure{Reweight}{$H^i,K^i,V_2^i$}\Comment{See [17] for implementation}
    \State{\textbf{return} $\hat{J},\hat{c} $}
    \EndProcedure
    \For{$(u,v)\in G^i[K^i\cup V^i_1]$, $u \neq v$}
    \If{$(u,v) \in K^i \times K^i$}
    \State{$J^{i+1}_{uv} += \hat{J}^{i+1}_{uv}$}
    \Else
    \State{$J^{i+1}_{uv} = {J}^{i+1}_{uv}$}
    \EndIf
    \EndFor
    \State{$V^{i+1} = \{K^i \cup V_1^i\}$}
    \State{$E^{i+1}=E(V_1^i)\cup E(K^i \times K^i)$}
    \State{$\mathcal{G}^{i} = (V^{i+1},E^{i+1},J^{i+1})$}
    \State{$G^{i+1} = \mathcal{G}^i$}
    \State{$c^{i} = c^i + \hat{c}$}
    \State{increment $i$}
    \EndWhile
    \State{\textbf{return} $\mathcal{G}^i,c^i$}
    \EndProcedure
    \Output{$\mathcal{{G}}^i,c^i$}
    \end{algorithmic}
\end{algorithm}

\section{Results}\label{sec:results}

We perform several numerical experiments to evaluate QEGD performance on several different graph families. All QAOA expectation values are obtained with the Qiskit Aer simulator with a COBYLA optimizer and 1,500 shots, except for the experiment in Sec.~\ref{expC}, where expectation values are calculated using Eq.~\eqref{eq:EXP1}. For all experiments, the objective maximum value used to calculate the approximation ratios is obtained by formulating MaxCut on the original graph as an LP and solving in Gurobipy. All graphs are obtained through Python NetworkX's $random\_regular\_graph()$ and $erdos\_renyi()$ functions. 

\subsection{Merge Threshold Effect}\label{sec:resA}

In this experiment, we empirically test for a baseline $\tau$ threshold which allows for relatively high accuracy merges while also not constricting the amount of merges that occur. Too high of a $\tau$ would limit severely the amount of merges which occur, and thus negating the purpose of the merge routine. Too low of a $\tau$ will allow for merges between vertices that are not highly correlated, which may result in a suboptimal solution. We ran 50 20-node $d$-regular graphs with $d \in \{3,4,5,6\}$,  varying $\tau \in \{0.1,0.2,0.3, 0.4\}$ for each $d$, and compared the results to GDOA on the same seeded, random regular graphs. For QEGD, our termination criteria is $M =5$. $C_{max}$ is obtained for each problem graph using Python's GurobiPy, and the approximation ratio is calculated for each problem instance. Each problem instance is solved 10 times for GDOA and QEGD, and their averages are taken as the final approximation ratio metric. 

 For $3-$ and $4-$regular graphs, the approximation ratio obtained after applying QEGD is approximately the same as the approximation ratio resulting from GDOA . However, for $5-$ regular graphs at $\tau = 0.2$, we see a $3\%$ increase in approximation ratio, and a $4.8\%$ increase for $6-$regular graphs when compared to GDOA. These experimental results are shown in Table~\ref{tab:merge-sweep}. The minimal differences in approximation ratios for 3- and 4-regular graph runs most likely indicate that very few merges, which can only occur when $|K| > 3$, actually triggered before the termination criteria was met. For the 5- and 6-regular graphs, it is plausible that more $|K| > 3$ merges occurred, and avoided the imperfect reweighting slack errors that GDOA incurred. It is also worth noting that the approximation peaked when $\tau = 0.3$ for the $3-$regular graphs, while it peaked when $\tau=0.2$ for the $4-$ and $5$-regular graphs. These results imply that $\tau = 0.2$ provides a favorable threshold for merge confidence, particularly on higher degree graphs.

 \subsection{Higher Order Merge Threshold Effect}
It then behooves us to see if the approximation ratio trends and comparisons between GDOA and QEGD continue into regular graphs with a greater number of vertices. We established an experiment which ran QEGD and GDOA on a dataset of twenty-five 100-node $d$-regular graphs, $d \in \{3,4,5,6\}$ for varying $\tau \in{0.1,0.2}$. The termination criteria for QEGD was $M = 10$, and all other parameters were kept the same as the previous experiment in Sec.~\ref{sec:resA}. Each problem instance was ran 10 times, and the approximation ratios were averaged for the final metric. These results are organized in Table~\ref{tab:merge-sweep2}. The data shows that GDOA and QEGD achieved similar approximation ratios for 3- and 4-regular graphs. For 5- and 6-regular graphs, the data shows that QEGD produces a substantial increase in approximation ratio over GDOA for both $\tau = 0.1$ and $\tau = 0.2$. This result may indicate that the GDOA reweighting algorithm may have substantial slack terms in the LP subroutine resulting in a suboptimal reweighting while there may be few erroneous $\tau =0.1$ merges in QEGD. We do note that the approximation ratios slightly decrease as $|V|$ grows, which may be a result of some vertices having increased degree, and thus more two-qubit gates in the QAOA subroutine. This pattern is consistent with the motivation for QEGD:  the improved accuracy of the QEGD algorithm over GDOA comes in the form of reduced error terms at each vertex merging (reweighting) step. 

To see the scalability of QEGD, we run the algorithm on 3-,4-,5-, and 6-regular random graphs with between $100$ and $800$ nodes. The approximation ratio is averaged over 30 graphs per configuration, and we record net node reduction, merge-trigger counts, and mean $|K|$-reduction per merge. The experiment is performed for QEGD at $\tau =0.1$ and $\tau = 0.2$, where $M$ for each experiment was selected as $\frac{|V|}{10}$. The data for 3 and 4 regular graphs are found in Table~\ref{tab:tau_stats_34reg}, and the 5 and 6 regular graph data is shown in Table~\ref{tab:tau_stats_56reg}. One quantifiable pattern is that $\tau = 0.2$ obtains higher approximation ratios than $\tau = 0.1$ nearly all across the board, showing an improvement of up to 6\%. This indicates that $\tau = 0.1$ is not an optimal merge threshold, even though it does produce larger net reductions and larger mean $\Delta|K|$ per merge than $\tau = 0.2$. Notably, $\tau = 0.2$ admits fewer merges, but the merges result in a higher approximation ratio. The results of this experiment show the limits of QEGD with regard to approximation ratio, vertex size, and graph degree. Specifically, QEGD approximation ratio does reduce much quicker with denser graphs because the inexact reweighting still occurs more often in more dense graphs than less dense graphs.

\subsection{Approximation ratio and wall-clock time comparison between QEGD and other Iterative Quantum Algorithms}\label{expC}
In this section, we evaluate QEGD performance (approximation ratio and wall-clock time) and compare it to  QAOA$^2$, DC-QAOA, and ParaQAOA, which all use a graph decomposition routine in conjunction with QAOA to solve graph-based optimization problems \cite{zhou2023qaoa, li2022large, huang2026paraqaoaefficientparalleldivideandconquer}.  
All experiments in this work are performed on sets of Erd\"{o}s-R\'{e}nyi (ER) random graphs with between 20 and 30 nodes. The Erd\"{o}s-R\'{e}nyi graphs have edge probabilities, $\rho$, between $0.1$ and $0.8$. The QEGD merge threshold value was set as $\tau = 0.2$ and $M = 7$ for all ER graphs. A single $p$ layer QAOA is utilized for all algorithms, and expectation values are calculated using Eq.~ \ref{eq:EXP1}.

\begin{figure}[h]
    \centering
    \includegraphics[width=01.0\linewidth]{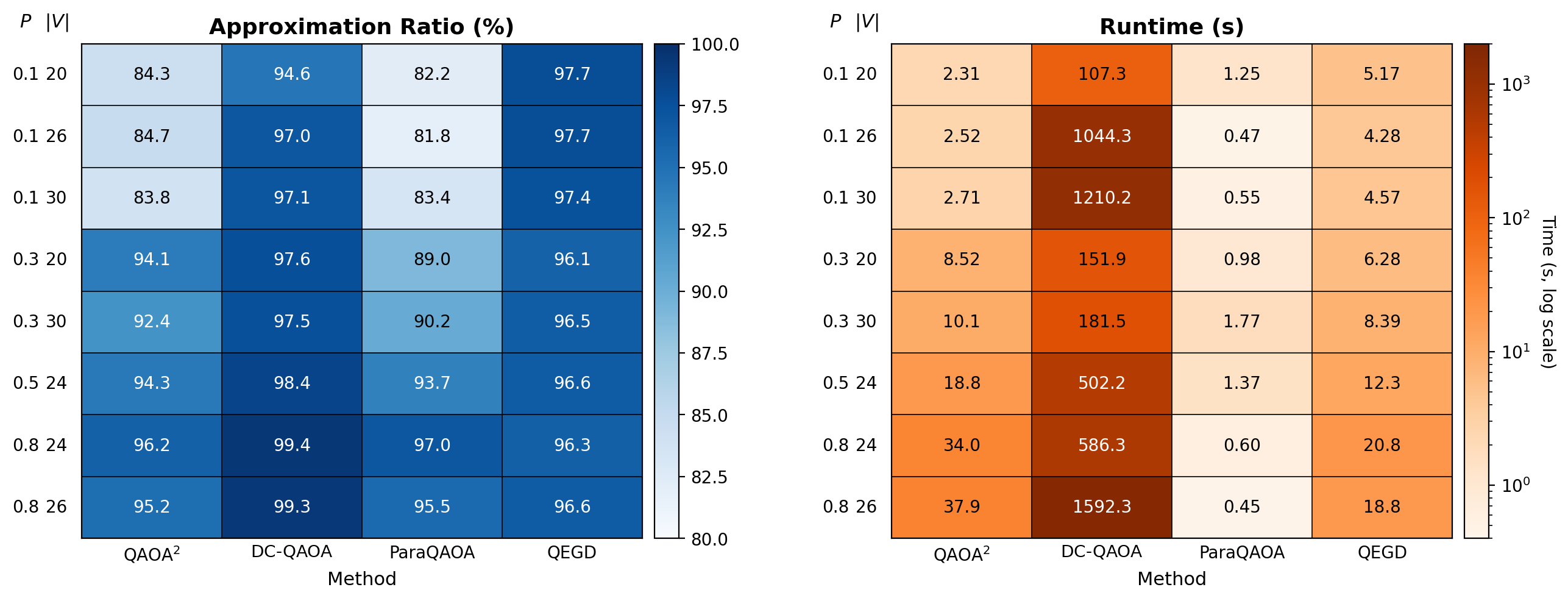}
    \caption{ Comparison between QAOA in QAOA (QAOA$^2$), Divide and Conquer QAOA (DC-QAOA), ParaQAOA, and our QEGD. a) heatmap of approximation ratios between algorithms b) total wall time. }
    \label{fig:hm}
\end{figure}

 We see that in sparse ($\rho=0.1$) graphs, QEGD achieves the highest approximation ratio of all four methods, slightly edging out DC-QAOA while also running much faster. ParaQAOA and QAOA$^2$ both trail substantially in AR at this density, but have lower wall-clock times. For denser graphs, ($\rho = 0.8$), DC-QAOA overtakes QEGD in approximation ratio, but its runtime grows to roughly $30$--$85$ times that of QEGD's. ParaQAOA remains the fastest method by a wide margin at every density but no longer trails in AR at high density, closing to within $1$--$4$ points of DC-QAOA and QEGD.
Overall, QEGD occupies a favorable region of the quality-time trade-off space. QEGD matches or exceeds the DC-QAOA approximation ratio at low density while avoiding its runtime blow-up. It also has a significantly higher approximation ratio than ParaQAOA at low-to-moderate density. The comparison is illustrated in Fig.~\ref{fig:hm}.

\subsection{Error Accumulation from Erroneous Merging}
However, we obviously still do accrue error from erroneous merges, e.g. merges that fix two nodes $x_i$ and $x_j$ to have the same value even though they do not have the same value in the optimal solution. After an analysis on a 100-node experiment for 3-,4-,5-, and 6- regular graphs and $\tau = 0.2$, we find that $\approx$8-11\% of merges are result in suboptimal node assignments. This metric was determined by running the algorithm, then once the final super-node assignments were found, working backward chronologically and counting the nodes whose assignment deviates from the optimal solution. As expected,  the assignment obtained by calculating QAOA correlations on subgraphs is local and thus is not always aligned with the global assignment. 
It is worth noting that $\tau$ was fixed in this experiment, so an open research direction is to determine if graph topology has an impact on which $\tau$ results in the fewest suboptimal merges.

\section{Conclusion}\label{sec:concs}
In this work, we introduce a Quantum Enhanced Graph Decomposition (QEGD) algorithm which utilizes QIRA design principles. The algorithm utilizes QAOA as a subroutine to inform a heuristic node merge where an expectation value greater than $\tau$ merges two nodes into a supernode. We compare this new algorithm against GDOA for 3-,4-,5-, and 6- regular graphs and find that QEGD algorithm results in 15\% higher approximation ratio than GDOA on average for 6-regular graphs, as well as a 10\% higher approximation ratio than GDOA on average for 5-regular graphs. Finally, we compared the runtime and approximation ratio to distributed QAOA algorithms, and found that QEGD has comparable approximation ratio to DC-QAOA and has a significantly shorter wall-clock time. It also has a higher approximation ratio than QAOA$^2$ and ParaQAOA, as well as comparable wall-clock time on the tested datasets. On sparse, connected, ER graphs with 20, 26, and 30 nodes at $\rho = 0.1$, QEGD achieves 97.7\%. 97.7\%. and 97.4\% approximation ratio, respectively. 

The results in this work were collected by numerically simulating QEGD on graphs with different vertex sizes, edge densities, and with different merge criteria $\tau$. Directions for future research include developing machine learning models to predict optimal values of $\tau$ and $M$ or using machine learning models to predict the QIRA or QAOA variant that results in the highest approximation ratio for a given problem instance. Additionally, the merge routine in this algorithm reduces the problem graph by calculating local QAOA correlations between vertices, which can result in suboptimal merges. Thus, an interesting research direction would be to develop a method that can determine when a local merge is suboptimal and correct the merge without solving the entire problem instance. Another interesting direction would be to determine if the suboptimal local merges are dependent on problem structure. Finally, MaxCut is an unconstrained combinatorial optimization problem, and thus every bitstring in the solution space is feasible. We note that this is not always true for constrained optimization problems. Thus, it would be interesting to determine if new graph reduction methods (aside from local merges) can be used to ensure that a quantum-inspired problem reduction scheme only results in feasible solutions.

\section*{Author contributions}
M.B. and L.P. contributed to the experiment design and developed the code. R.H. contributed to experiment design. All authors wrote, read, edited, and approved the final manuscript.

\section*{Acknowledgments}
M. Brunet acknowledges NSF award DGE-2152168. The funder played no role in study design, data collection, analysis and interpretation of data, or the writing of this manuscript.

\section*{AI use}
The author used Claude Fable 5.1 to generate some code in output formatting, testing, and Tikz, as well as to generate Fig.~\ref{fig:mergeGraphic}. The author takes sole responsibility for correctness of the final manuscript

\section*{Code, data availability, and reproducibility}
 The QEGD code and experiment graphs throughout this work can be found at \cite{qe_repo}. The code for QAOA$^2$, DC-QAOA, and ParaQAOA was obtained from  \cite{qaoa2_repo}, \cite{dc_repo}, and \cite{para_repo} respectively. 
  All code was executed locally on a base 3.10 GHz AMD Ryzen 5 5500U processor, with up to 6 cores and 12 threads available, 16 GB of RAM, Ubuntu 24.04, Python 3.12, Qiskit 1.3.2, PennyLane 0.40, GurobiPy 12.0.3, NumPy 1.26, and NetworkX 2.8.8. 
 
\section*{Competing interests}
All authors declare no financial or non-financial competing interests. 

\appendix

\section{Data Tables}
\begin{table}[h]
  \centering
  \caption{Effect of the $|K|>3$ supernode-merge
  step. GDOA is the
  common $\tau$-independent baseline; each
  MERGE($\tau$) is compared against it. Each approximation ratio (AR) is calculated as the average of 10 runs for each of the 50 graphs.}
  \label{tab:merge-sweep}
  \begin{tabular}{llc}
  \toprule
  Benchmark & Algorithm & AR \\
  \midrule
  \multirow{5}{*}{\shortstack[l]{3-regular\\20
  nodes\\50 graphs}}
   & GDOA            & 0.9745 \\
   & QEGD($\tau{=}0.1$) & 0.9686 \\
   & QEGD($\tau{=}0.2$) & 0.9698 \\
   & QEGD($\tau{=}0.3$) & 0.9775 \\
   & QEGD($\tau{=}0.4$) & 0.9753 \\
  \midrule
  \multirow{5}{*}{\shortstack[l]{4-regular\\20
  nodes\\50 graphs}}
   & GDOA            & 0.9632 \\
   & QEGD($\tau{=}0.1$) & 0.9507 \\
   & QEGD($\tau{=}0.2$) & 0.9687 \\
   & QEGD($\tau{=}0.3$) & 0.9695 \\
   & QEGD($\tau{=}0.4$) & 0.9795 \\
  \midrule
  \multirow{5}{*}{\shortstack[l]{5-regular\\20
  nodes\\50 graphs}}
   & GDOA            & 0.9410 \\
   & QEGD($\tau{=}0.1$) & 0.9315 \\
   & QEGD($\tau{=}0.2$) & 0.9690 \\
   & QEGD($\tau{=}0.3$) & 0.9593 \\
   & QEGD($\tau{=}0.4$) & 0.9478 \\
  \bottomrule
  \midrule
\multirow{5}{*}{\shortstack[l]{6-regular\\20 nodes\\50 graphs}}
 & GDOA            & 0.9267 \\
 & QEGD($\tau{=}0.1$) & 0.9393 \\
 & QEGD($\tau{=}0.2$) & 0.9720 \\
 & QEGD($\tau{=}0.3$) & 0.9602 \\
 & QEGD($\tau{=}0.4$) & 0.9555 \\
\bottomrule

  \end{tabular}
  \end{table}

\begin{table}[t]
  \centering
  \caption{Comparison of original Graph Decomposition vs. QEGD}
  \label{tab:merge-sweep2}
  \begin{tabular}{llc}
  \toprule
  Benchmark & Algorithm & $AR$\\
  \midrule
  \multirow{5}{*}{\shortstack[l]{3-regular\\ 100
  nodes\\25 graphs}}
   & GDOA           & 0.9539 \\\\
   & QEGD($\tau{=}0.1$) & 0.9412 \\
   & QEGD($\tau{=}0.2$) & 0.9714 \\\\
  \midrule
  \multirow{5}{*}{\shortstack[l]{4-regular\\100
  nodes\\25 graphs}}
   & GDOA            & 0.9242 \\\\
   & QEGD($\tau{=}0.1$) & 0.9312 \\
   & QEGD($\tau{=}0.2$) & 0.9501 \\
   &  \\
  \midrule
  \multirow{5}{*}{\shortstack[l]{5-regular\\100
  nodes\\25 graphs}}
   & GDOA          & 0.8309 \\\\
   & QEGD($\tau{=}0.1$) & 0.9109 \\
   & QEGD($\tau{=}0.2$) & 0.9301 \\\\
   \midrule
  \multirow{5}{*}{\shortstack[l]{6-regular\\100
  nodes\\25 graphs}}
   & GDOA          & 0.8009 \\\\
   & QEGD($\tau{=}0.1$) & 0.9316 \\
   & QEGD($\tau{=}0.2$) & 0.95015 \\\\
  \bottomrule

  \end{tabular}
  \end{table}

   \begin{table}[p]
  \centering
  \small
  \begin{tabular}{llcccccc}
  \toprule
  \textbf{Graph Family} & \textbf{$|V|$} & \textbf{$\tau$} & \textbf{AR} & \textbf{Net Reduction} & \textbf{Merge Triggers} & \textbf{$\overline{\Delta |K|}$} & \textbf{$\overline{\Delta |K|} > 0$} \\
  \midrule
  3-Regular & 100 & 0.1 & 0.9623 & 41.50  & 29.85  & 1.689 & 1.904 \\
            &     & 0.2 & 0.9691 & 40.25  & 28.75  & 1.340 & 1.713 \\ \addlinespace
            & 200 & 0.1 & 0.9410 & 89.25  & 70.10  & 1.797 & 1.949 \\
            &     & 0.2 & 0.9584 & 89.19  & 68.9   & 1.507 & 1.802 \\ \addlinespace
            & 300 & 0.1 & 0.9356 & 135.35 & 102.65 & 1.861 & 1.990 \\
            &     & 0.2 & 0.9553 & 133.8  & 100.5  & 1.568 & 1.822 \\ \addlinespace
            & 400 & 0.1 & 0.9321 & 181.45 & 150.80 & 1.906 & 2.021 \\
            &     & 0.2 & 0.9531 & 178.6  & 143.7  & 1.611 & 1.838 \\ \addlinespace
            & 500 & 0.1 & 0.9296 & 227.10 & 176.55 & 1.938 & 2.045 \\
            &     & 0.2 & 0.9514 & 223.1  & 170.6  & 1.643 & 1.851 \\ \addlinespace
            & 600 & 0.1 & 0.9277 & 272.40 & 213.10 & 1.962 & 2.064 \\
            &     & 0.2 & 0.9501 & 267.4  & 207.3  & 1.667 & 1.861 \\ \addlinespace
            & 700 & 0.1 & 0.9262 & 317.45 & 259.45 & 1.980 & 2.079 \\
            &     & 0.2 & 0.9490 & 311.5  & 253.8  & 1.686 & 1.870 \\ \addlinespace
            & 800 & 0.1 & 0.9250 & 362.25 & 285.65 & 1.994 & 2.091 \\
            &     & 0.2 & 0.9481 & 355.4  & 280.2  & 1.701 & 1.877 \\
  \midrule
  4-Regular & 100 & 0.1 & 0.9323 & 55.40  & 24.40  & 1.905 & 2.284 \\
            &     & 0.2 & 0.9392 & 53.14  & 21.55  & 1.554 & 2.382 \\ \addlinespace
            & 200 & 0.1 & 0.9107 & 119.15 & 49.45  & 2.042 & 2.451 \\
            &     & 0.2 & 0.9189 & 115.65 & 46.90  & 1.718 & 2.491 \\ \addlinespace
            & 300 & 0.1 & 0.9041 & 181.20 & 74.15  & 2.121 & 2.549 \\
            &     & 0.2 & 0.9127 & 176.4  & 70.9   & 1.812 & 2.552 \\ \addlinespace
            & 400 & 0.1 & 0.9001 & 242.55 & 98.60  & 2.174 & 2.612 \\
            &     & 0.2 & 0.9090 & 236.3  & 94.6   & 1.874 & 2.594 \\ \addlinespace
            & 500 & 0.1 & 0.8973 & 303.40 & 122.85 & 2.212 & 2.658 \\
            &     & 0.2 & 0.9063 & 295.6  & 118.1  & 1.919 & 2.625 \\ \addlinespace
            & 600 & 0.1 & 0.8952 & 363.85 & 146.95 & 2.241 & 2.694 \\
            &     & 0.2 & 0.9043 & 354.5  & 141.4  & 1.954 & 2.649 \\ \addlinespace
            & 700 & 0.1 & 0.8935 & 423.95 & 170.90 & 2.264 & 2.722 \\
            &     & 0.2 & 0.9028 & 413.1  & 164.6  & 1.982 & 2.668 \\ \addlinespace
            & 800 & 0.1 & 0.8921 & 483.75 & 194.75 & 2.283 & 2.745 \\
            &     & 0.2 & 0.9015 & 471.4  & 187.7  & 2.005 & 2.684 \\
  \bottomrule
  \end{tabular}
  \caption{3- and 4-Regular stats averaged over 30 graphs per size and $\tau$ value. From left to right: $\tau$ merge threshold, average approximation ratio, average total merges, average $|K|$ reduction per merge, average non-zero $|K|$ reduction.}
  \label{tab:tau_stats_34reg}
  \end{table}

  \begin{table}[p]
  \centering
  \small
  \begin{tabular}{llcccccc}
  \toprule
  \textbf{Graph Family} & \textbf{$|V|$} & \textbf{$\tau$} & \textbf{AR} & \textbf{Net Reduction} & \textbf{Merge Triggers} & \textbf{$\overline{\Delta |K|}$} & \textbf{$\overline{\Delta |K|} > 0$} \\
  \midrule
  5-Regular& 100 & 0.1 & 0.9209 & 56.25  & 29.90  & 1.687 & 1.943 \\
            &     & 0.2 & 0.9391 & 54.50  & 28.95  & 1.285 & 1.916 \\ \addlinespace
            & 200 & 0.1 & 0.9101 & 115.4  & 58.15  & 1.740 & 2.043 \\
            &     & 0.2 & 0.9148 & 113.2  & 56.84  & 1.407 & 2.085 \\ \addlinespace
            & 300 & 0.1 & 0.9095 & 173.7  & 86.05  & 1.776 & 2.104 \\
            &     & 0.2 & 0.9106 & 171.0  & 84.35  & 1.478 & 2.180 \\ \addlinespace
            & 400 & 0.1 & 0.9090 & 231.5  & 113.70 & 1.803 & 2.147 \\
            &     & 0.2 & 0.9081 & 228.2  & 111.60 & 1.529 & 2.246 \\ \addlinespace
            & 500 & 0.1 & 0.9016 & 288.9  & 141.15 & 1.824 & 2.180 \\
            &     & 0.2 & 0.9034 & 284.9  & 138.65 & 1.567 & 2.295 \\ \addlinespace
            & 600 & 0.1 & 0.8983 & 346.0  & 168.45 & 1.841 & 2.206 \\\\
          & 700 & 0.1 & 0.8910 & 402.8  & 195.60 & 1.855 & 2.227 \\
          &     & 0.2 & 0.9041 & 397.3  & 192.20 & 1.623 & 2.365 \\ \addlinespace
          & 800 & 0.1 & 0.8778 & 459.4  & 222.65 & 1.867 & 2.245 \\
          &     & 0.2 & 0.8933 & 453.1  & 218.75 & 1.644 & 2.391 \\
\midrule
6-Regular & 100 & 0.1 & 0.9321 & 58.80  & 32.90  & 1.682 & 1.996 \\
          &     & 0.2 & 0.9415 & 52.2   & 30.2   & 1.136 & 1.840 \\ \addlinespace
          & 200 & 0.1 & 0.9124 & 120.9  & 64.35  & 1.742 & 2.097 \\
          &     & 0.2 & 0.9401 & 107.9  & 59.45  & 1.258 & 1.947 \\ \addlinespace
          & 300 & 0.1 & 0.9058 & 182.4  & 95.40  & 1.782 & 2.160 \\
          &     & 0.2 & 0.9282 & 162.8  & 88.30  & 1.331 & 2.011 \\ \addlinespace
          & 400 & 0.1 & 0.8918 & 243.4  & 126.10 & 1.812 & 2.205 \\
          &     & 0.2 & 0.9047 & 217.2  & 116.85 & 1.383 & 2.056 \\ \addlinespace
          & 500 & 0.1 & 0.8791 & 304.0  & 156.55 & 1.835 & 2.239 \\
          &     & 0.2 & 0.9011 & 271.2  & 145.20 & 1.422 & 2.090 \\ \addlinespace
          & 600 & 0.1 & 0.8670 & 364.2  & 186.80 & 1.853 & 2.266 \\
          &     & 0.2 & 0.8812 & 324.9  & 173.35 & 1.454 & 2.117 \\ \addlinespace
          & 700 & 0.1 & 0.8554 & 424.1  & 216.85 & 1.868 & 2.288 \\
          &     & 0.2 & 0.8687 & 378.3  & 201.35 & 1.480 & 2.139 \\ \addlinespace
          & 800 & 0.1 & 0.8341 & 483.7  & 246.75 & 1.881 & 2.306 \\
          &     & 0.2 & 0.8574 & 431.5  & 229.20 & 1.502 & 2.157 \\
\bottomrule
\end{tabular}
\caption{5- and 6-regular stats averaged over 30 graphs per size and $\tau$ value.}
\label{tab:tau_stats_56reg}
\end{table}
\clearpage
\bibliographystyle{ieeetr}
\bibliography{mainFinal}
\end{document}